\documentclass[conference]{IEEEtran}
\IEEEoverridecommandlockouts
\usepackage{cite}
\usepackage{textcomp}
\usepackage{xcolor}
\usepackage{fancyhdr}
\usepackage{amsmath,amssymb,amsfonts}
\usepackage{algorithmic}
\usepackage{graphicx}
\usepackage{textcomp}
\usepackage{xcolor}
\usepackage{svg}
\usepackage{multirow}
\usepackage{indentfirst}
\usepackage{array}
\usepackage{tabularray}
\usepackage{pgfplots}
\usepackage[linesnumbered]{algorithm2e}
\usepackage{calc}
\usepackage{subcaption}
\usepackage{enumitem}
\usepackage{setspace}
\usepackage{makecell}
\usepackage{booktabs} 
\usepackage{array}
\RestyleAlgo{ruled}
\usepackage{amssymb} 
\usepackage{pifont}  
\usepackage{adjustbox}
\usepackage{subfig}
\usepackage{marvosym} 

\SetKwInOut{KwIn}{Input}
\SetKwInOut{KwOut}{Output}

\usepackage{fancyhdr} 
\usepackage{lipsum}   
\usepackage{fancyhdr} 
\def\BibTeX{{\rm B\kern-.05em{\sc i\kern-.025em b}\kern-.08em
    T\kern-.1667em\lower.7ex\hbox{E}\kern-.125emX}}

\begin{document}

\title{
ERASER: \underline{E}fficient \underline{R}TL F\underline{A}ult \underline{S}imulation Framework with Trimmed \underline{E}xecution \underline{R}edundancy
}
\author{\IEEEauthorblockN{
Jiaping Tang\textsuperscript{*}\textsuperscript{\dag}\textsuperscript{\ddag},
Jianan Mu\textsuperscript{*}\textsuperscript{\dag}\textsuperscript{\Letter}, 
Silin Liu\textsuperscript{*}\textsuperscript{\dag}\textsuperscript{\ddag},
Zizhen Liu\textsuperscript{*}\textsuperscript{\dag}, 
Feng Gu\textsuperscript{*}\textsuperscript{\dag}\textsuperscript{\ddag},
Xinyu Zhang\textsuperscript{*}\textsuperscript{\dag}\textsuperscript{\ddag},
Leyan Wang\textsuperscript{*}\textsuperscript{\dag}\textsuperscript{\ddag},
\\
Shengwen Liang\textsuperscript{*}\textsuperscript{\dag},
Jing Ye\textsuperscript{*}\textsuperscript{\dag}\textsuperscript{\ddag},
Huawei Li\textsuperscript{*}\textsuperscript{\dag}\textsuperscript{\ddag}\textsuperscript{\Letter},
Xiaowei Li\textsuperscript{*}\textsuperscript{\dag}
}
\IEEEauthorblockA{
\textsuperscript{*}\textit{State Key Lab of Processors, Institute of Computing Technology, Chinese Academy of Sciences, Beijing, China} \\
\textsuperscript{\dag}\textit{University of Chinese Academy of Sciences, Beijing, China}\\
\textsuperscript{\ddag}\textit{CASTEST Co., Ltd., Beijing, China}\\
\{tangjiaping22s, mujianan, liusilin23s, liuzizhen, gufeng22s, zhangxinyu23s, wangleyan24s, \\ liangshenwen, yejing, lihuawei, lxw\}@ict.ac.cn}
}

\maketitle \pagestyle{empty}
\thispagestyle{fancy} 

\begin{abstract}


As intelligent computing devices increasingly integrate into human life, ensuring the functional safety of the corresponding electronic chips becomes more critical. 
A key metric for functional safety is achieving a sufficient fault coverage. 
To meet this requirement, extensive time-consuming fault simulation of the RTL code is necessary during the chip design phase.
The main overhead in RTL fault simulation comes from simulating behavioral nodes (always blocks). 
Due to the limited fault propagation capacity, fault simulation results often match the good simulation results for many behavioral nodes.
A key strategy for accelerating RTL fault simulation is the identification and elimination of redundant simulations.
Existing methods detect redundant executions by examining whether the fault inputs to each RTL node are consistent with the good inputs.
However, we observe that this input comparison mechanism overlooks a significant amount of implicit redundant execution: although the fault inputs differ from the good inputs, the node's execution results remain unchanged. 
Our experiments reveal that this overlooked redundant execution constitutes nearly half of the total execution overhead of behavioral nodes, becoming a significant bottleneck in current RTL fault simulation.
The underlying reason for this overlooked redundancy is that, in these cases, the true execution paths within the behavioral nodes are not affected by the changes in input values. In this work, we propose a behavior-level redundancy detection algorithm that focuses on the true execution paths. Building on the elimination of redundant executions, we further developed an efficient RTL fault simulation framework, Eraser.
Experimental results show that compared to commercial tools, under the same fault coverage, our framework achieves a 3.9 $\times$ improvement in simulation performance on average.

\end{abstract}


\section{Introduction}

\begin{figure}[htbp]
    \centering
    \setlength{\abovecaptionskip}{-0.00cm}
    \setlength{\belowcaptionskip}{-0.00cm}
    \includegraphics[width=\linewidth]{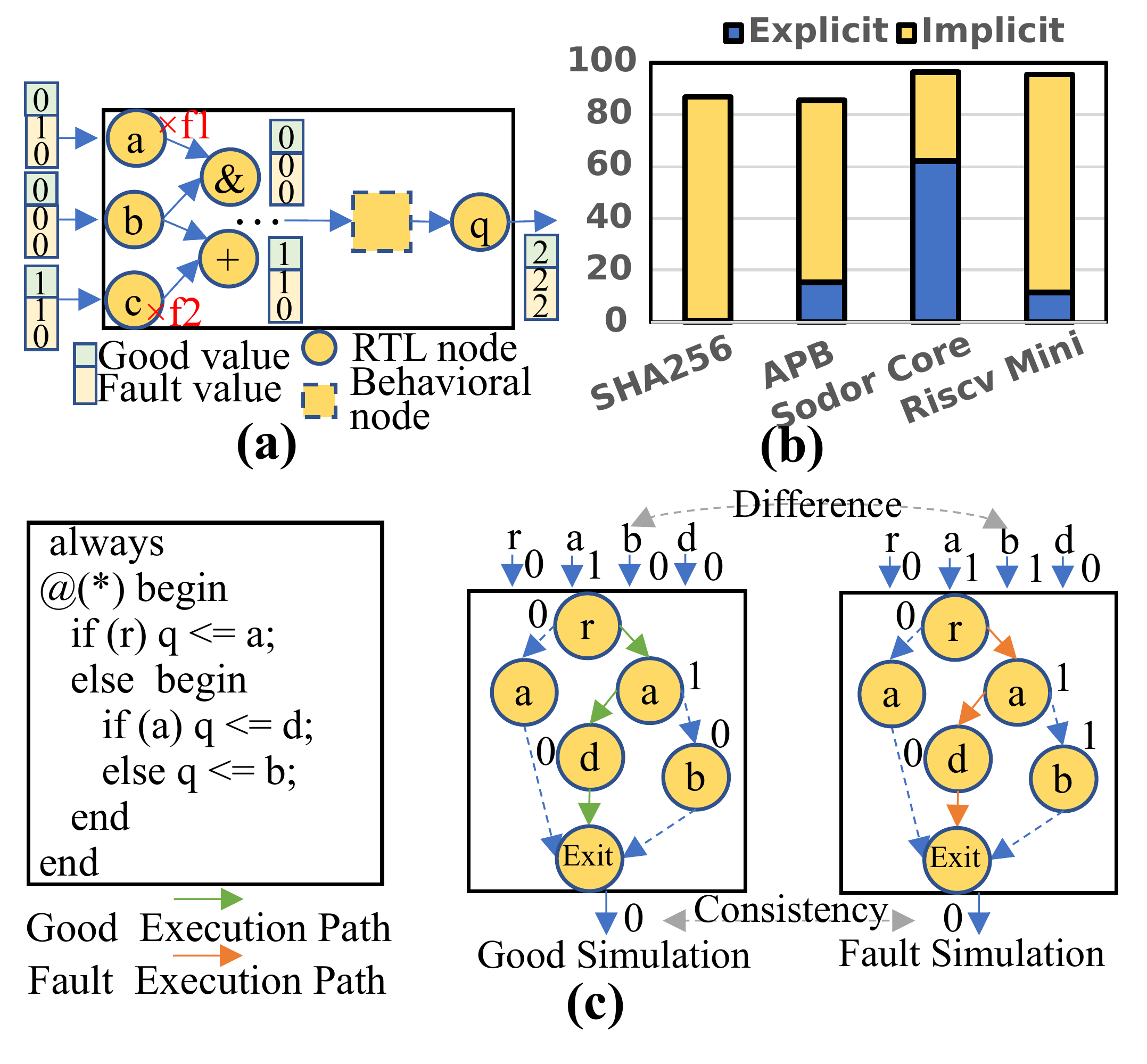}
    \caption{
        (a) The execution redundancy in RTL fault simulation. 
        (b) The ratio of explicit and implicit redundancy. 
        Explicit redundancy occurs when the good and fault inputs of a node are identical, while implicit redundancy occurs when the fault inputs differ from the good, yet retain the same outputs.
        (c) Example of implicit redundancy.
    }
    \label{intro}
\end{figure}

%
With the growing integration of intelligent computing systems like autonomous vehicles and smart robotics into human life, the functional safety of electronic chips is receiving heightened attention. The ISO 26262 standard \cite{ISO} is introduced to define specific functional safety requirements for automotive-grade chips, such as microcontroller unit (MCU), application-specific integrated circuit (ASIC), and system-on-chip (SoC).
According to ISO 26262, ensuring functional safety requires high fault coverage in chip design, necessitating extensive fault simulations and significant time costs~\cite{ISS_1}.
Compared to the high computational cost of gate-level simulation and the insufficient accuracy of instruction-level simulation, RTL-level fault simulation has become the mainstream approach for functional safety verification.
However, RTL fault simulations are also computationally intensive, often becoming a bottleneck in real-world development cycles\cite{long_em, long_2}.
Therefore, accelerating RTL fault simulation is urgently needed.

The RTL design consists of behavioral nodes (codes of always blocks) and RTL nodes (logical connections linking these behavioral nodes)\cite{iverilog}, which shows in Fig .\ref{intro}(a). 
Fault simulation for RTL code involves stimulating behavioral and RTL nodes with both good and faulty inputs.
Based on our measurements, shown in Section \ref{ablation_section}, the simulation time cost of behavioral nodes accounts for 60\% of the total cost. 
However, due to the limited fault propagation capabilities in most cases, the propagation of both faulty and good values tends to be consistent, leading to identical simulation results across a large number of nodes. 
Therefore, fault simulations that produce the same results as good simulations at a given node are redundant. 
We illustrate these redundancies in Fig.~\ref{intro}(a). 
Our experiments indicate that such redundant executions account for approximately 90\% of the behavioral node simulation time. 
Consequently, the key to improving fault simulation performance is to eliminate these redundant executions. 
Existing RTL fault simulation frameworks \cite{mozart, multi_cs} detect the inputs to each node, and for a behavioral node, if its faulty inputs match the good inputs, the corresponding fault simulation is skipped.

However, this approach of directly checking node inputs fails to fully eliminate the redundant execution of behavioral nodes. 
We find that beyond the explicit redundant executions where the inputs between good and fault simulations are identical, a significant amount of implicit redundancies for behavioral nodes remains unsettled.
We illustrate such a case in Fig.~\ref{intro}(c): although the fault inputs differ from the good inputs, the final output remains the same. 
Since the fault inputs are different from the good inputs, this redundant execution is overlooked by the existing method.
We test the proportion of these implicit redundant executions across several circuits. 
The results shown in Fig.~\ref{intro}(b) indicate that these implicit redundancies account for almost half of the total behavioral node executions in these cases.
Therefore, the challenge in improving the performance of RTL fault simulation is addressing implicit redundancies of behavioral nodes.

To this end, we propose Eraser, an efficient batched RTL fault simulation framework to eliminate redundant executions. 
First, we develop a redundancy detection algorithm based on execution paths at run-time. 
This algorithm identifies implicit redundancy by checking the consistency of execution paths and data dependencies.
Next, we integrate this algorithm with explicit redundancy detection, which is based on input comparison at behavioral nodes, to create a simulation framework that thoroughly eliminates behavioral node redundancies. 
By removing the redundancies more effectively, our framework achieves significant performance improvement compared to existing RTL fault simulation methods.
The main contributions are summarized as follows.
\begin{itemize}
    \item We identify significant implicit redundancies of behavioral nodes in RTL fault simulation and propose a redundancy detection algorithm based on execution paths at runtime to eliminate them.
    \item We propose the Eraser framework, which aims to exploit implicit redundancy detection while effectively managing both good and faulty behavioral executions as well as RTL node operations.
    \item The experimental results show that compared to the state-of-the-art commercial simulator and an open-source fault simulator, our simulator achieves an average performance acceleration of 3.9$\times$ and 5.9$\times$, respectively.
\end{itemize}

\section{Background}\label{section:background} 

\subsection{RTL code and fault simulation}
\begin{figure}[htbp]
\setlength{\abovecaptionskip}{-0.00cm}
\setlength{\belowcaptionskip}{-0.3cm}
\centerline{
\includegraphics[width=\linewidth]{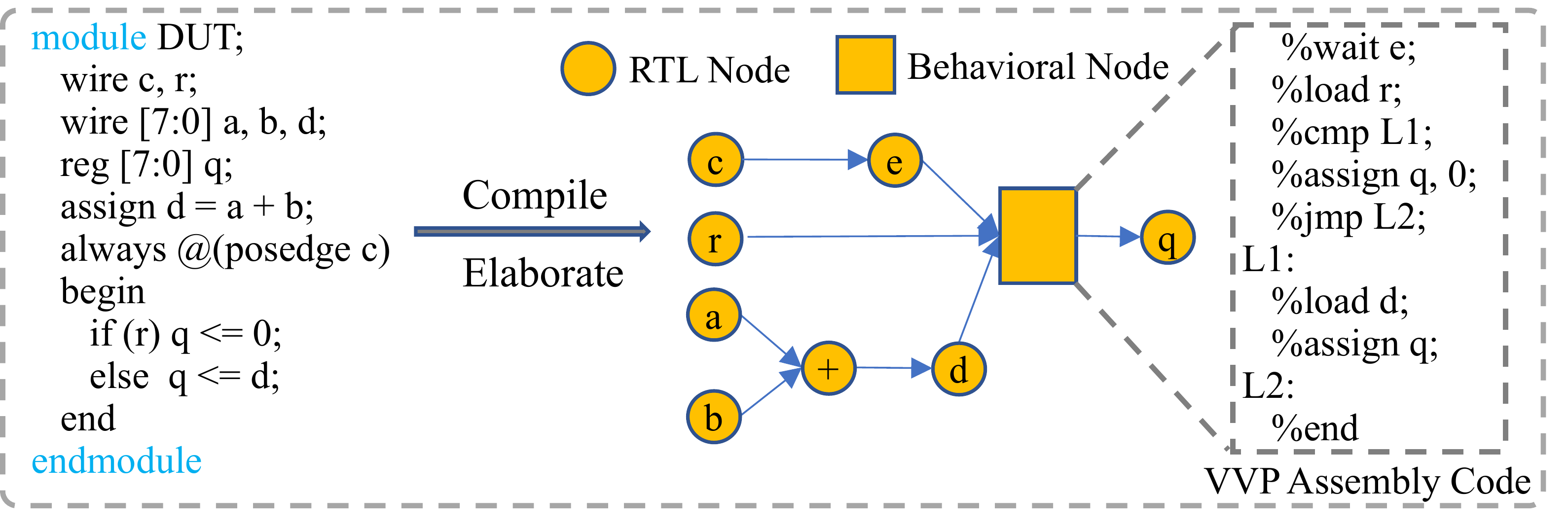}}
\caption{A RTL code and the internal representation\cite{iverilog}. 
}
\label{iverilog}
\end{figure}

The RTL code can be represented as a directed graph\cite{iverilog}, as shown in Fig.~\ref{iverilog}. The directed graph termed the RTL Graph, represents the signal connections between variables, arithmetic and logic operations, and behavioral code in the RTL design. We refer to the nodes composed of behavioral codes as behavioral nodes, and other nodes as RTL nodes.

Concurrent is a batch fault simulation algorithm widely used in gate-level fault simulation, which we have extended to the RTL level. To facilitate understanding in the following sections, several key terms need to be clarified.
\begin{itemize}
    \item \textbf{Good gate and bad gate.} A good gate exists in the fault-free network and maintains a list of bad gates. A bad gate exists in the faulty network and is used to store any differences arising from faults.
    \item \textbf{Visible and invisible bad gates.} If the output value of a bad gate is identical to the good gate, then the bad gate is considered invisible. Conversely, the bad gate is visible.
    \item \textbf{Good and faulty behavioral code.} A good behavioral code is activated by a good gate, with the execution path referred to as the good execution path. Conversely, a faulty behavioral code is activated by a bad gate, with the execution path referred to as the faulty execution path.
\end{itemize}

\subsection{Related work}

Despite considerable progress, RTL fault simulation methods still face key challenges. 
Most approaches only support single fault simulation \cite{FI_3 ,mefisto ,verify, HL_1, ISS_1, ISS_2, FSim_1, FSim_2}, and fail to address path convergence effectively, leading to redundant computations and suboptimal performance. 
While some efforts have explored concurrent fault simulation at the RTL level \cite{mozart, multi_cs}, these solutions remain incomplete, especially in dealing with implicit redundancies in behavioral code, which contribute significantly to computational overhead.

Techniques such as RTL code modification \cite{FI_3 ,mefisto ,verify} and using built-in commands in simulators \cite{vcs, iverilog} have been proposed, but these methods either require extensive manual intervention or rely heavily on specific tools. 
Higher-level fault simulation \cite{HL_1, ISS_1, ISS_2} can improve efficiency but often sacrifices accuracy. Recent efforts to enhance Verilator \cite{FSim_1, FSim_2} still lack support for batch fault simulation, limiting scalability.
Multilevel batch fault simulators \cite{mozart, multi_cs} offer better performance but do not fully address redundancies in behavioral nodes. 
Therefore, eliminating these redundancies remains a crucial opportunity for improving RTL fault simulation efficiency.


\section{Motivation of Eraser}\label{section:motivation}

\begin{figure}[htbp]
\setlength{\abovecaptionskip}{-0.00cm}
\setlength{\belowcaptionskip}{-0.00cm}
\centerline{
\includegraphics[width=\linewidth]{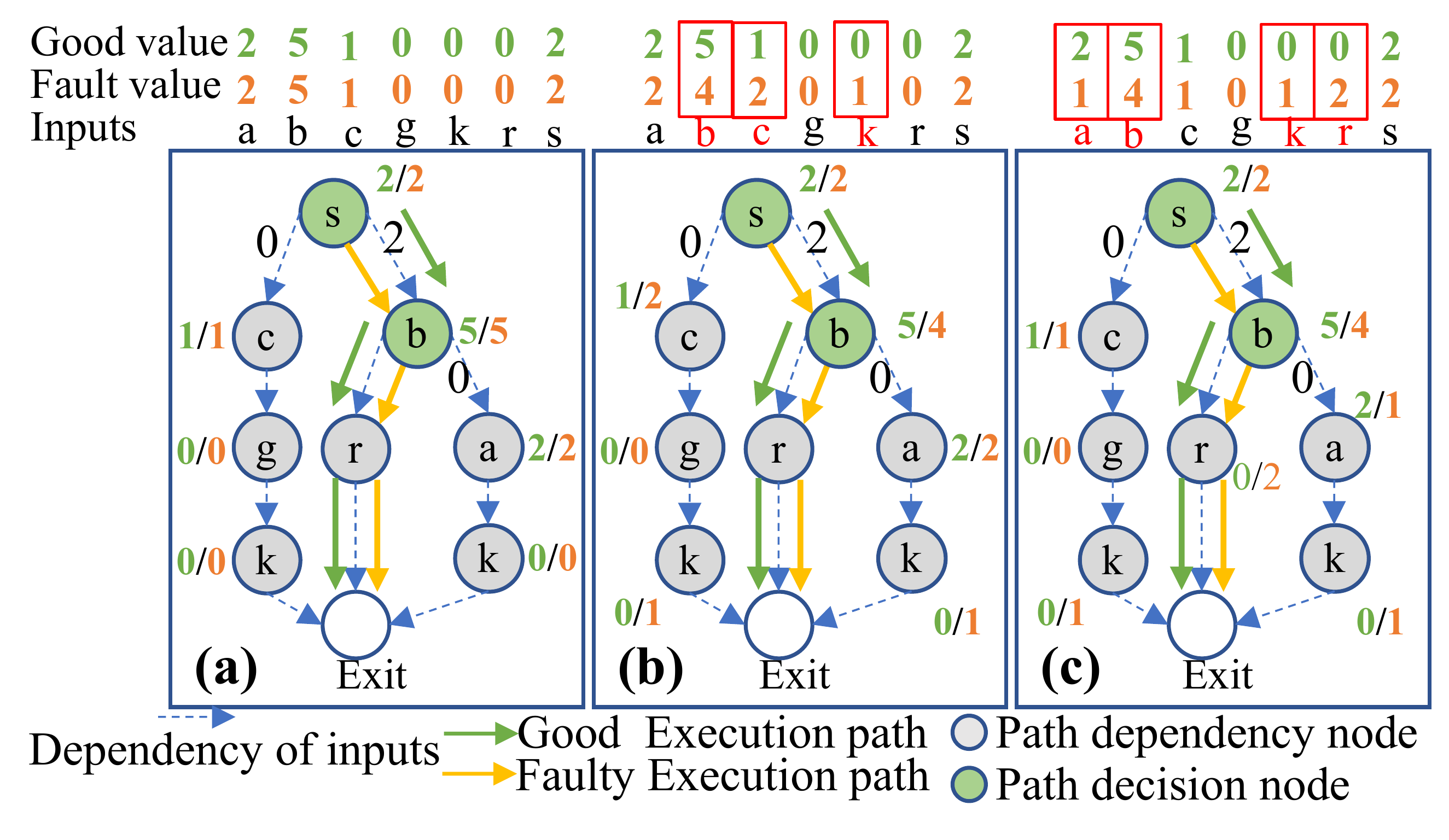}
}
\caption{Comparison between impliciy and explicit redundancy.}
\label{fig:motivation}
\end{figure}

To analyze the core mechanisms of implicit redundancy, we compare the implicit redundancy and explicit redundancy in Fig.~\ref{fig:motivation}, where the box illustrates the internal execution path under different input values of a behavior code.
The nodes are categorized into two types: path decision nodes (green) and data dependency nodes (gray). Path decision nodes represent branch statements in RTL, while data dependency nodes indicate data dependencies along the path. Given a set of inputs, a specific execution path is selected, and data propagation along this path produces a result.

Fig.~\ref{fig:motivation}(a) illustrates an explicit redundancy node, where the inputs remain unchanged, leading to the same output.
Fig.~\ref{fig:motivation}(b) represents an implicit redundancy. In this case, the fault inputs for signals \textit{b}, \textit{c}, and \textit{k} differ from the good ones. Specifically, signal \textit{b} is a path decision node and the value is changed, 
but the execution path remains unchanged.
Meanwhile, these input changes do not appear in the data dependencies of the actual execution path, resulting in identical simulation outcomes for both good and faulty behavioral codes. This indicates execution redundancy at the behavioral node.
In Fig.~\ref{fig:motivation}(c), the changes in the fault do not affect the execution path, but the value of signal \textit{r} influences the data dependencies on the actual path,  causing a different result. Therefore, this cannot be identified as a redundant execution.
From the above comparison, it can be concluded that detecting implicit redundancy is to determine whether the differences between the fault and good values affect the choice of execution path and the data dependencies on this path.

\section{Eraser Framework with redundant node detection}\label{section:framework}

\begin{figure}[htbp]
\setlength{\abovecaptionskip}{-0.00cm}
\setlength{\belowcaptionskip}{-0.00cm}
\centerline{
\includegraphics[width=\linewidth]{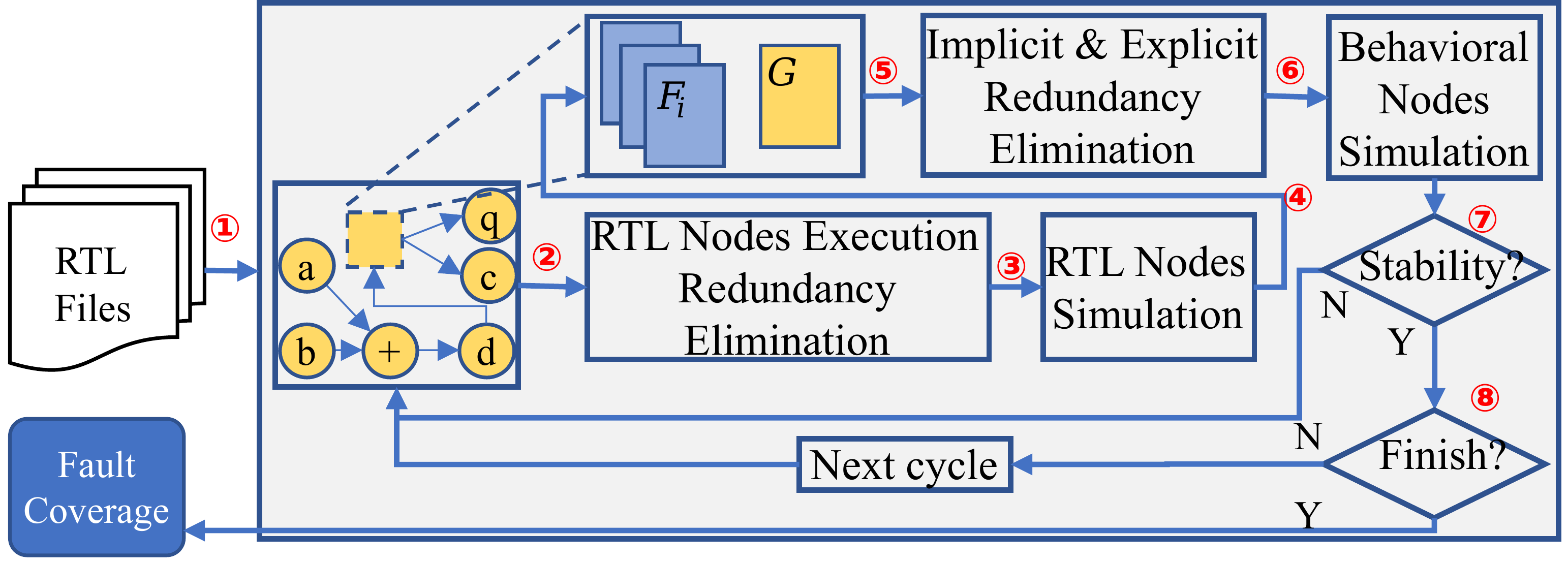}
}
\caption{Framework of Eraser.}
\label{overview_fig}
\end{figure}

The framework takes the RTL code as input and ultimately outputs the fault coverage of the circuit. The framework consists of eight steps.
Step 1 is to compile and elaborate the RTL code to generate an RTL graph, which includes RTL nodes and behavioral nodes.
The concurrent fault simulation is used to reduce execution redundancy on the RTL nodes in step 2 and 3. Some good and faulty behavioral codes will be activated by RTL nodes after RTL nodes simulation in step 4. Faulty behavioral codes will be skipped if redundancy detection methods determine them to be redundant in step 5. Otherwise, they will be executed in step 6.
It is worth noting that ERASER introduces redundancy detection based on the actual execution flow to identify implicit redundant that previous methods have overlooked. A detailed explanation of this detection method is provided in Section \ref{behavioral}. Additionally, we have reproduced the explicit redundancy detection methods for behavior and RTL nodes from existing approaches, as explained in Section \ref{behavioral_explicit} and \ref{netlist}.
Steps 2 to 6 are iterated based on whether the RTL nodes or behavioral nodes have reached a stable state, or if the entire simulation has been completed, otherwise, output the fault coverage.

%


\subsection{Implicit Redundancy Elimination and Fault Simulation on Behavioral Nodes}\label{behavioral}

\begin{figure*}[htbp]
\setlength{\abovecaptionskip}{-0.00cm}
\setlength{\belowcaptionskip}{-0.00cm}
\centerline{
\includegraphics[width=0.95\linewidth]{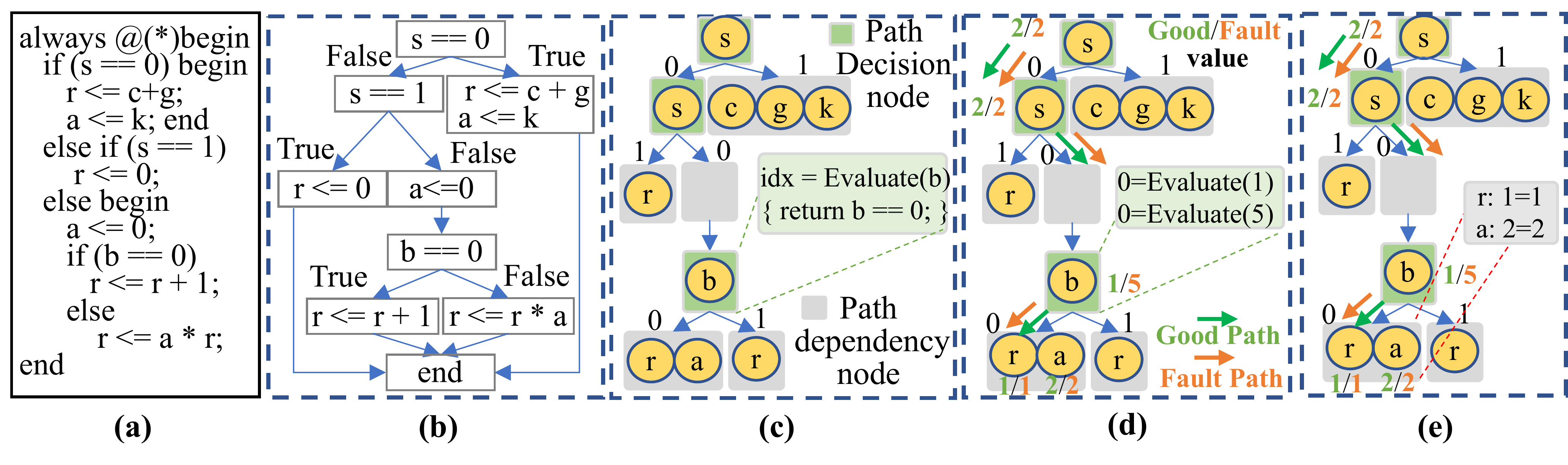}
}
\caption{An example of how to determine execution path redundancy.}
\label{elimination}
\end{figure*}

To efficiently eliminate redundant execution in behavioral nodes, we propose Algorithm~\ref{alg_execution}, an execution flow redundancy elimination approach. Algorithm~\ref{alg_execution} takes behavior nodes, and fault id as input. 
By using the good execution path as a reference and monitoring the path decision nodes and path dependency nodes, it detects the redundant execution of faulty behavioral codes where the fault input differs from the good input but the output remains consistent with the good behavioral codes.
We describe the Algorithm~\ref{alg_execution} in the context of a circuit case, as shown in Fig.~\ref{elimination}.


\begin{algorithm}
\caption{Implicit Redundancy Elimination} 
\label{alg_execution}
\KwIn{ Behavior\_code: A fault-free behavioral code; fault\_id: Specified the faulty behavioral code}
\KwOut{The faulty behavioral code is redundant or not }
CFG $\gets$ build\_control\_flow\_graph(Behavior\_code)\;
VDG $\gets$ build\_visibility\_dependency\_graph(CFG)\;
cur $\gets$ the entry node of VDG\; 
\While{ cur not null} { 
    \If{cur is path decision node}{
        next\_good\_node $\gets$ Evaluate(good gates)\;
        next\_fault\_node $\gets$ Evaluate(bad gates)\;
        \If{ next\_good\_node $\neq$ next\_fault\_node}{
            \Return false\;
        }
    } 
    \If {cur is path dependency node}{
        \ForEach{signal $\in$ cur} {
            \If{IsVisible(signal, fault\_id)} {
                \Return false\;
            }
        }
    }
    cur $\gets$ next\_node;
}
\Return true\;
\end{algorithm} 

\textbf{Preprocess (Line 1 in Algorithm \ref{alg_execution})}: To trace the execution path of behavioral code, we first partition the code, with each partition representing a potential execution segment where no branching occurs. We utilize control flow analysis techniques to achieve this partitioning, converting the behavioral code into a control flow graph (CFG). As shown in Fig.~\ref{elimination}(b), this represents the control flow graph for Fig.~\ref{elimination}(a).

\textbf{Visibility dependency graph construction (Line 2 in Algorithm \ref{alg_execution})}:
Determining the dominance relationships of various input signals under different values is important to identify which input signals truly impact the execution results of the behavioral code and which do not.
To obtain the dependency relationships between different input signals across various execution paths, we extend the CFG to construct a visibility dependency graph of input signals.
Fig.~\ref{elimination}(c) illustrates the visibility dependency graph derived from Fig.~\ref{elimination}(b). As shown, the structure of the dependency graph mirrors that of the CFG, with each node storing the input signals that are read by the corresponding execution segment.
Besides, we can simplify the visibility dependency graph by removing empty nodes.

This graph effectively captures the varying execution paths of the behavioral code under different inputs. Moreover, by focusing on a specific execution path, we can identify which input data at any given time influences the execution results of behavioral codes. 
For instance, as shown in Fig.~\ref{elimination}(c), when the input signal \textit{s} assumes the value of 1, the signals \textit{c}, \textit{g} and \textit{k} are dominated, while \textit{a} and \textit{r} are not. Thus, only the values of \textit{c}, \textit{g}, and \textit{k} will affect the execution result, while the signals \textit{a} and \textit{r} will not.
In the visibility dependency graph, the nodes are categorized into two types: path decision nodes and path dependency nodes. Path decision nodes determine the actual execution path at a given time. Each decision node contains an \textit{Evaluate} function extracted from CFG, which assesses the current signal values to determine the appropriate sub-path. 
In contrast, path dependency nodes store the input signals on which a sub-path relies during execution. These signals affect the final result of the execution path.

\textbf{Check on path decision point (Line 5-11 in Algorithm \ref{alg_execution})}:
A faulty behavioral code is considered redundant only if its execution path exactly matches that of the good behavioral code, and the input signals along the path are identical. To determine whether a faulty behavioral code is redundant, the first step is to check if its execution path aligns with that of the good behavioral code. 
Therefore, we need to analysis the path decision nodes.
The next sub-path for the behavioral code can be determined by the output of the \textit{Evaluate} function at the path decision node. For both the good behavioral code and the faulty behavioral code, we first retrieve their respective input signals and then invoke the \textit{Evaluate} function to determine the subsequent execution paths (Lines 6-7 of the algorithm), as shown in Fig.~\ref{elimination}(d) determining the sub-path for node \textit{b}. 
%
If the paths at the path decision node are not consistent between the two, we conclude that the execution of the faulty behavioral code is non-redundant. Otherwise, further analysis of the subsequent paths is required.


\textbf{Check on path dependency point (Line 12-18 in Algorithm \ref{alg_execution})}:
Under the condition that the execution paths are consistent, the agreement of the path dependency nodes affects whether the execution results of the good behavioral code and the faulty behavioral code are aligned. Therefore, we need to assess the consistency of the path dependency nodes.
Since each path dependency node stores all the input signals that the behavioral code relies on for that sub-path, we must traverse these signals and check each for equality, as shown in Fig.~\ref{elimination}(e) checking the visibility for both input signal \textit{a} and \textit{r}. 
If any input signals on this sub-path differ, we conclude that the execution results of the faulty behavioral code and the good behavioral code are inconsistent, indicating that the faulty behavioral code is non-redundant. Otherwise, further analysis of the subsequent paths is necessary.

\subsection{Explicit execution redundancy detection}\label{behavioral_explicit}

Concurrent fault simulation only simulates activated events in the fault-free network plus any differences that arise due to faults\cite{multi_cs}, which has been used to reduce explicit redundancy. So, we extend the concurrent fault simulation algorithm from gate level to RTL to eliminate the explicit redundancy existing in behavioral nodes. 


\subsection{Simulation on the RTL Nodes}\label{netlist}
We enhance the concurrent fault simulation capabilities for all RTL nodes (logic nodes, arithmetic nodes, and others) in Iverilog. Due to Iverilog's inherent event scheduling strategy, where certain nodes are evaluated immediately upon receiving new values while others are deferred\cite{iverilog}, issues, such as fake events, arise during the extension. Those issues may undermine the accuracy of fault simulation results. In our framework, we resolve this problem by adjusting the evaluation and scheduling relationships of the RTL nodes.


Fake events refer to the premature evaluation of bad gates on event nodes in Iverilog, due to the impact of good events, resulting in erroneous activation of waiting behavioral codes. 
 Because of good events are sent to their successors before bad events, which causes the bad gate of event nodes to mistakenly detect an event when receiving good values, resulting in incorrect activation of waiting behavioral codes. 
 Specifically, because good events are sent to successors before bad events, the bad gate of event nodes mistakenly detects an event when receiving good values, while the actual arrival of bad events may generate different events, resulting in incorrect activation of waiting behavioral codes.
 To avoid abnormal activation caused by fake events, we postpone the evaluation of all event nodes until all blocking events\cite{ieee_verilog} have been processed. 

\section{Evaluation}\label{section:eveulation}

\subsection{Experimental Settings}
We implement Eraser in C++ and the evaluation environment is shown in Table \ref{test:env}. 
For comparison, we chose three simulators that support RTL fault simulation. The first one is a state-of-the-art commercial simulator, Z01X\cite{z01x}.
The second one is an open-source fault simulator\cite{FSim_1} which is based on Verilator\cite{verilator}, we called VFsim for convenience.
Besides, we implemented fault simulation in Icarus Verilog\cite{iverilog} using the \textit{force} command, which we refer to as IFsim.

\begin{table}[h]
\centering
\setlength{\abovecaptionskip}{0cm}
\caption{Evaluation Environment}
\label{test:env}
\resizebox{\linewidth}{!}{
\begin{tabular}{|c|c|}
\hline
\textbf{Filed} & \textbf{Value } \\[0.5pt]
\hline
\hline
CPU & Intel(R) Xeon(R) Platinum 8260 CPU @ 2.40GHz \\
OS & Red Hat Enterprise Linux Server 7.9(Maipo) \\
Compiler & gcc 11.1.0, -O3 \\
\hline
\multirow{4}{*}{Simulator} & 
Z01X T-2022.06-SP2(Z01X) \\ &
\cite{FSim_1} 2021 based on Verilator(VFsim) \\ & 
Iverilog 12(IFsim)\\ 
\hline
\end{tabular}
}
\end{table}

\textbf{Benchmark and testbench.} 
Table \ref{benchmark} presents the benchmarks used in this paper, covering arithmetic cores\cite{ALU, FPU}, encryption cores\cite{sha256}, communication controllers\cite{OpenCores}, RISCV CPUs\cite{ucb_bar, picorv32}, MIPS CPUs\cite{mips}, and accelerators\cite{sha256_accel, conv_1}. 
In our experiments, the stimuli used are either sourced from test benches provided by the design developers or written by us based on the function of benchmarks if test benches are not provided. The simulation cycles for each design are also shown in Table \ref{benchmark}.

\begin{table}[h]
\centering
\setlength{\abovecaptionskip}{0cm}
\caption{Benchmark Information}

\label{benchmark}
\resizebox{.95\linewidth}{!}{
\begin{tabular}{|c|c|c|c|c|c|}
\hline
\multirow{2}{*}{\textbf{Benchmark}} &
\multirow{2}{*}{\textbf{\#Stimulus}} &
\multirow{2}{*}{\textbf{\#Cells}} &
\multirow{2}{*}{\textbf{\#Faults}} &
\multicolumn{2}{c|}{\textbf{Fault coverage(\%)}} 
 \\
\cline{5-6}
& & & & \textbf{Eraser} & \textbf{Z01X} \\

\hline
\hline
ALU (64) & 1.5k & 19996 & 1182 & 95.69 & 95.69  \\
\hline
FPU (32) & 9k & 8875 & 1256 & 99.04 & 99.04  \\
\hline
SHA256\_HV$^{\ast}$ & 2.6k & 8677 & 660 & 99.85 & 99.85 \\
\hline
APB & 1.2k & 7051 & 98 & 91.84 & 91.84  \\
\hline
Sodor Core & 3k & 16943 & 1252 & 81.07 & 81.07 \\
\hline
RISCV Mini &  6k & 9087 &  526 & 27.97 & 27.97 \\
\hline
PicoRV32 & 4k  & 17488 & 1040 & 32.79 & 32.79  \\
\hline
Conv\_acc & 4k & 39812 & 1032 & 79.75 & 79.75 \\
\hline
SHA256\_C2V$^{\star}$ & 4k  & 9716 & 2174 & 99.31 & 99.31  \\
\hline
MIPS CPU & 1.2k  & 15000 & 1346 & 44.40 & 44.40  \\
\hline
\multicolumn{6}{l}{\textbf{\#Cells}: Number of cells reported by the Yosys tool\cite{Yosys}. } \\
\multicolumn{6}{l}{\textbf{$\ast$ SHA256\_HV}: the handwritten Verilog of SHA256. } \\
\multicolumn{6}{l}{\textbf{$\star$ SHA256\_C2V}: the Verilog generated by Chisel of SHA256. }
\end{tabular}
}
\end{table}

\textbf{Fault list and fault coverage.} 
We generate stuck-at faults for wires and regs in designs and set observation points at all output ports.
An observation is triggered to determine whether any faults have been detected when good events occur at observation points.
As shown in Table \ref{benchmark}, the Eraser fault coverage is consistent with that of the commercial tool Z01X across all benchmarks, demonstrating that the accuracy of the fault simulation framework proposed in this paper.

\subsection{Evaluation on Benchmarks and Comparisons}

\begin{figure*}[htbp]
\setlength{\abovecaptionskip}{-0.00cm}
\setlength{\belowcaptionskip}{-0.00cm}
\centerline{
\includegraphics[width=\linewidth]{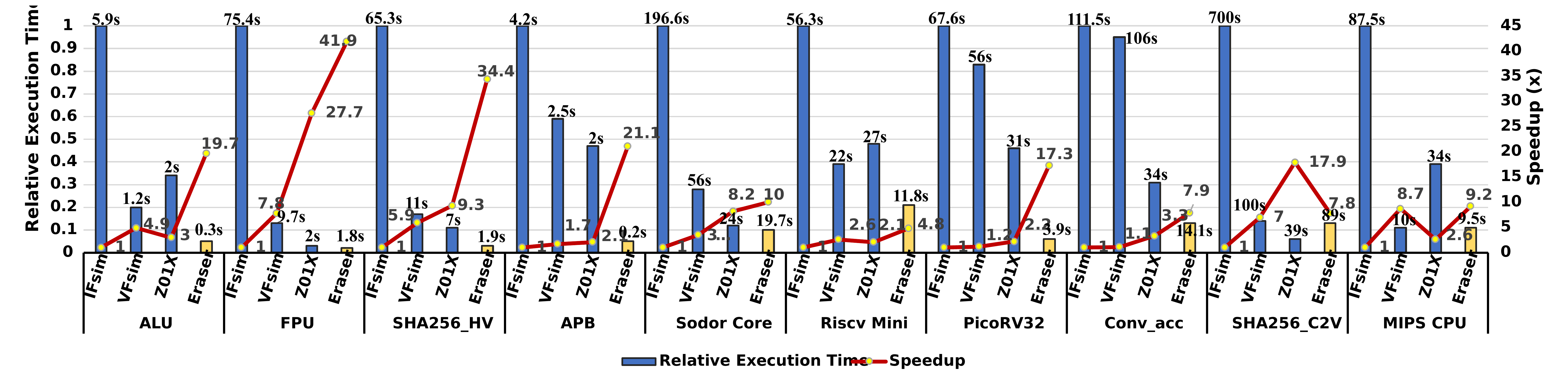}
}
\caption{ Performance comparison of various RTL fault simulators. The \textit{IFsim} is the baseline.}
\label{performance_cmp}
\end{figure*}

We conducted performance tests on all the aforementioned testbenches and compared them with mainstream commercial tools and the latest open-source simulators. 
Notably, Eraser achieved the same fault coverage as commercial tools, as shown in TABLE \ref{benchmark}, demonstrating the correctness of our algorithm. 
The experimental results are shown in Fig.~\ref{performance_cmp}, which includes the algorithm's execution time and corresponding speedup ratios. 
As observed, Eraser achieved the best performance across all benchmark circuits except for SHA256\_C2V. 
For example, in the arithmetic unit ALU, the execution times of IFsim, VFsim, Z01X, and Eraser are 5.9s, 1.2s, 2s, and 0.3s, respectively. 
Compared to the baseline IFsim, they achieved speedup ratios of 4.9$\times$, 3.0$\times$, and 19.7$\times$, respectively. 
For the floating-point unit FPU, VFsim, Z01X, and Eraser achieved speedup ratios of 7.8$\times$, 27.7$\times$, and 41.9$\times$ compared to IFsim.

Eraser outperforms the commercial tool Z01X on most benchmarks, especially on the APB benchmark for a speedup of 10.0$\times$ and on the ALU benchmark for a speedup of 6.7$\times$. 
However, in the case of the SHA256\_C2V benchmark, the performance of Eraser is inferior to that of Z01X. 
The reason can be attributed to two aspects.
First, the execution time of the behavioral codes in SHA256\_C2V accounts for only 1\% of the total execution time, which significantly limits the effectiveness of optimizations for redundant execution of behavioral nodes.
Second, the commercial Z01X employs various engineering optimizations for node execution that Eraser has not yet utilized. We plan to incorporate these optimizations in future work to further enhance Eraser's performance.
On average, Eraser achieves a 3.9$\times$ speedup compared to commercial tool and a 5.9$\times$ speedup compared to the latest open-source tools.

\subsection{\textbf{Ablation study}}\label{ablation_section}

\begin{figure}[htbp]
\setlength{\abovecaptionskip}{-0.00cm}
\setlength{\belowcaptionskip}{-0.00cm}
\centerline{
\includegraphics[width=\linewidth]{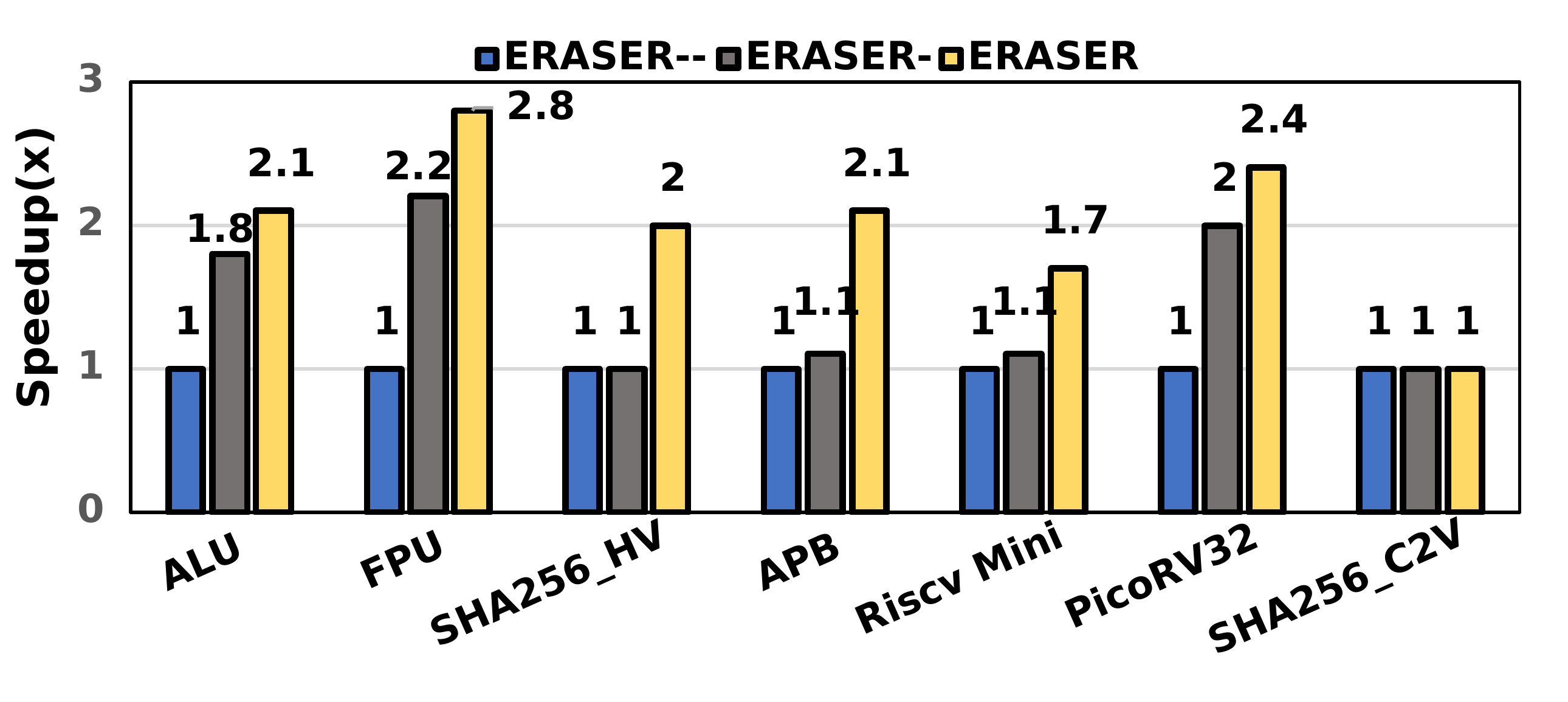}
}
\caption{ Ablation Study on Redundancy Elimination: Eraser\text{-}\text{-} (No Redundancy Elimination), Eraser- (Eliminates Explicit Redundancy), Eraser (Eliminates Both Explicit and Implicit Redundancy)}
\label{speedup}
\end{figure}



We conducted an ablation study to demonstrate the acceleration effect of implicit redundancy elimination on RTL fault simulation. We implement another two versions of Eraser: Eraser\text{-}\text{-}, which does not detect any redundancy; Eraser-, which detects only explicit redundancy. Aside from redundancy detection, they have the same implementation as Eraser. In Fig.~\ref{speedup}, we present the speedup across various circuits. The comparison shows that Eraser achieves significant speedup over both Eraser\text{-}\text{-} and Eraser- for most circuits.
For example, for the FPU circuit, Eraser achieves a 2.8$\times$ speedup over Eraser\text{-}\text{-} and a 1.3$\times$ speedup over Eraser-. Similarly, for the SHA256\_HV implementation, Eraser achieves a 2.0$\times$ speedup over Eraser\text{-}\text{-} and a 2.0$\times$ speedup over Eraser-. 
However, for SHA256\_HV, Eraser shows similar performance with Eraser- and Eraser--. This is because, compared to the Verilog version, the Chisel implementation contains many RTL nodes, resulting in very few implicit and explicit redundancies in behavioral nodes.


\begin{table}[h]
\centering
\setlength{\abovecaptionskip}{0cm}
\caption{Proportion of Redundant Behavioral Node Executions}
\label{execution_cnt}
\resizebox{\linewidth}{!}{
\begin{tabular}{|c|c|c|c|c|c|}
\hline
\multirow{2}{*}{\textbf{Benchmark}} & \textbf{Time for} & \textbf{\#Total BN} &
\multirow{2}{*}{\textbf{\#Elimination}} &
\multirow{2}{*}{\textbf{Explicit(\%)}} &
\multirow{2}{*}{\textbf{Implicit(\%)}} \\
 & \textbf{BN(\%)} & \textbf{Execution} &  &  &  \\
\hline
\hline
ALU 
 & 57
 & 339592 
 & 324714 
 & 82 
 & 14\\
\hline
FPU 
 & 70
 & 1891740 
 & 1793457  
 & 81
 & 14 \\
\hline
SHA256\_HV 
 & 70
 & 992540 
 & 862612 
 & 1
 & 86\\
\hline
APB 
 & 74
 & 211000 
 & 180650 
 & 15
 & 70\\
\hline
RISCV Mini 
 & 53
 & 2779987
 & 2650970
 & 11
 & 84\\
\hline
PicoRV32 
 & 61
 & 5701568
 & 5650319
 & 86
 & 13 \\
\hline
SHA256\_C2V 
 & 1
 & 834539
 & 634533
 & 49
 & 27 \\
\hline
\textbf{Average}  & \multicolumn{3}{c|}{ - } & \textbf{ 46 } & \textbf{ 44 } \\
\hline
\end{tabular}
}
\caption*{
}
\end{table}


Table~\ref{execution_cnt} further illustrates the proportion of redundant executions in these circuits. We list the proportion of runtime for behavioral nodes, the total number of behavioral node executions (without eliminating any redundant executions), the total number of redundant executions, and portions of explicit and implicit redundancies.
As shown in Table~\ref{execution_cnt}, in circuits such as SHA256\_HV, APB, and RISCV Mini, implicit redundancies account for more than 70\% of the redundancy behavioral node executions. Correspondingly, Eraser demonstrates significant performance improvements over Eraser- and Eraser\text{-}\text{-} in these circuits. Meanwhile, implicit redundancies in the PicoRV32 only account for 13\%, which explains the limited performance gain of Eraser over Eraser- in Fig.~\ref{speedup}.
Additionally, in SHA256\_C2V, the overhead of behavioral node simulation accounts for only 1\% of the total, making the optimization of behavioral node execution less impactful. As a result, Eraser- and Eraser show performance similar to Eraser\text{-}\text{-}. 
On average, across these circuits, the proportions of implicit and explicit redundancies are both around 45\%.

The above analysis demonstrates that implicit redundancy constitutes a significant part of the execution in RTL fault simulations. By eliminating it, Eraser achieves noticeable performance improvements.


\section{Conclusion}\label{section:conclusion}
In this work, we identify implicit redundancy in the simulation of behavioral nodes in RTL fault simulation and propose a redundancy detection algorithm based on execution paths at runtime and data dependencies to eliminate it. We further implement an efficient RTL fault simulation framework that trimmed the redundancy. The experimental results show that compared to the commercial simulator and an open-source fault simulator, our simulator achieves an average acceleration of 3.9$\times$ and 5.9$\times$, respectively.

\section*{Acknowledgment}
This paper is supported in part by the National Natural Science Foundation of China (NSFC) under grant No. (92473203, 92373206), and in part by the Chinese Academy of Sciences under grant No. XDB0660100.
The corresponding authors are Jianan Mu and Huawei Li.

\vspace{12pt}

\end{document}